\documentclass[aps,preprint]{revtex4}
\usepackage{float}
\usepackage[dvips]{graphicx}
\usepackage{amssymb}
\usepackage{amsmath}
\usepackage{color}

\usepackage[normalem]{ulem}
\usepackage{gensymb}
\newcommand{\soutr}{\bgroup\markoverwith{\textcolor{red}{\rule[.5ex]{2pt}{1pt}}}\ULon}
\usepackage{natbib}
\begin{document}
	
\draft

\title{Coupled unidirectional chaotic microwave graphs}
\author{Omer Farooq, Afshin Akhshani, Micha{\l} {\L}awniczak, Ma{\l}gorzata Bia{\l}ous, and Leszek Sirko}
\address{Institute of Physics, Polish Academy of Sciences, Aleja Lotnik\'{o}w 32/46, 02-668 Warszawa, Poland\\
}
\date{\today}

\begin{abstract}
We investigate experimentally the undirected open microwave network $\Gamma $ with internal absorption composed of two coupled directed halves, unidirectional networks $\Gamma_{+} $ and $\Gamma_{-} $, corresponding to two possible directions of motion on their edges.  The two-port scattering matrix of the network $\Gamma$ is measured and the spectral statistics and the elastic enhancement factor of the network are evaluated. The comparison of the number of experimental resonances with the theoretical one predicted by the Weyl's law shows that within the experimental resolution the resonances are doubly degenerate. This conclusion was also corroborated by the numerical calculations. Though the network is characterized by the time reversal symmetry the missing level spectral statistics and the elastic enhancement factor are rather close to the Gaussian unitary ensemble predictions in random matrix theory. We used numerical calculations for the open non-dissipative quantum graph possessing the same structure as the microwave network $\Gamma$ to investigate the doublet structures in the spectrum which otherwise would not be experimentally resolved. We show that the doublet size distribution is close to the Poisson distribution.
\end{abstract}

\pacs{03.65.Nk,05.45.Mt}
\bigskip
\maketitle

\section{Introduction}

From mathematical point of view a quantum graph is a one-dimensional complex system with the Laplace operator $L(\Gamma) = -\frac{d^2}{dx^2}$  defined in the Hilbert space of square integrable functions \cite{Wigner1951,Haake2018,Weidenmuller2009,Berkolaiko2013,Kurasov2024}. The concept of quantum graphs was introduced by Pauling \cite{Pauling1936} to model organic molecules. Later, quantum graphs were applied in modelling and studying a large variety of different systems and theories such as quantum chaos, dynamical system theory, photonics crystals, superconductivity theory, microelectronics, etc \cite{Kowal1990, Imry1996, Rohde2011, Elster2015}. 

A quantum graph consists of one-dimensional edges $e_i$ which are connected at the vertices $v_i$. The propagation of a wave along an edge of the graph  is described by the one-dimensional Schr\"odinger equation. The boundary conditions are implemented on the wave functions entering and leaving the vertices. Commonly the Neumann (N) and Dirichlet (D) vertex boundary conditions are applied. The Neumann boundary condition imposes the continuity of waves propagating in the edges meeting at the vertex $v_i$ and vanishing of the sum of outgoing derivatives at $v_i$. The Dirichlet boundary condition demands vanishing of the waves at the vertex.
According to Bohigas-Giannoni-Schmit conjecture \cite{Bohigas1984,Guhr1998,Haake2018,Heusler2007,Mehta1990,Hul2004,Lawniczak2008} the spectral properties of quantum systems underlying classically chaotic dynamics can be modelled by appropriate Gaussian ensembles of the random matrix theory (RMT). In this approach three main symmetry classes are distinguished: the Gaussian orthogonal ensemble (GOE) and the Gaussian symplectic ensemble (GSE) with time-reversal invariance ($\mathcal{T}$-invariance), characterized respectively by the symmetry indices $\beta=1$ and $\beta=4$, and the Gaussian unitary ensemble (GUE) with broken time-reversal invariance, $\beta=2$. The Gaussian unitary ensemble characterizes chaotic systems with any spin, while the Gaussian orthogonal and symplectic ensembles describe quantum and wave-dynamical chaos in systems with integer and half-integer spins, respectively.

The experimental studies of complex quantum systems are in general very complicated and challenging. This problem, for a wide class of such systems, has been effectively resolved with the help of microwave networks. The one-to-one equivalence of the stationary Schr\"odinger equation describing quantum graphs and the telegraph equation describing microwave networks allows to simulate quantum graphs through the use of microwave networks \cite{Hul2004,Lawniczak2010}.

A unique versatility of microwave networks as wave simulators stems from their being the only systems which allow for simulation of quantum graphs whose properties are described by all three symmetry classes GOE \cite{Hul2004,Lawniczak2008,Hul2009,Hul2012,Lawniczak2012,Sirko2016,Dietz2017,Lawniczak2019,Lawniczak2020}, GUE \cite{Hul2004,Lawniczak2010,Lawniczak2011,Lawniczak2019b,Bialous2016,Lawniczak2020a} and GSE \cite{Stockmann2016,Lu2020,Lawniczak2023} in the framework of RMT. The GSE systems can only be experimentally investigated using microwave networks.

The other complex quantum systems can be simulated by microwave flat billiards \cite{Dietz2010,Yeh2013,Zheng2006,Stockmann1990,Sridhar1994,Sirko1997,Bauch1998,Hlushchuk2000,Blumel2001,Hlushchuk2001,
	Hlushchuk2001b,Dhar2003,Savytskyy2004,HemmadyPRL2005,Hul2005,Hul2005c,Dietz2015,Dietz2019a,Bialous2019} and atoms excited in strong microwave fields \cite{Jensen1991,Sirko1991,Bellerman1992,Buchleitner1993,SirkoPRL1993,Bayfield1995,Sirko1995,Sirko1996,Bayfield1985,Sirko2001,Galagher2016}.

Quantum graphs with GUE properties can be simulated experimentally by microwave networks with microwave circulators \cite{Lawniczak2010,Lawniczak2020a}. 
Recently, Akila and Gutkin \cite{Akila2015} have theoretically and numerically considered an undirected quantum graph $\Gamma$ composed of two unidirectional ones $\Gamma_{+} $ and $\Gamma_{-} $ in which the nearest-neighbor spacing distribution of the eigenvalues shows close to GUE statistics. An experimental realization of a single unidirectional graph has been recently presented in Ref. \cite{Che2022}. The direction of the wave propagating through the unidirectional network was controlled by applying microwave hybrid couplers and isolators \cite{Che2022}. Though the paper was mainly focused on the spectral statistics of the unidirectional  network some other characteristics of the network such as, e.g., correlation functions and the distribution of the reflection amplitude were also analyzed.

In this article we present the results of an experimental study of the coupled unidirectional systems $\Gamma_{+} $ and $\Gamma_{-} $ corresponding to the networks with two opposite directions of wave motion on their edges which together form the undirected microwave network $\Gamma $. The network $\Gamma $ is open and characterized by internal absorption. The two-port scattering matrices of different realizations of the network are measured to evaluate the spectral statistics, the reflection coefficient, the imaginary part of the Wigner reaction matrix, and the elastic enhancement factor of the network.

The comparison of the number of experimentally observed resonances with the theoretical one predicted by the Weyl's law shows that approximately only half of the resonances have been experimentally identified. Because the graphs  $\Gamma_+$ and $\Gamma_-$, for symmetry reasons, are closely doubly degenerate, the resonances within the spectral resolution of the measurements should be doubly degenerate. This conclusion was also corroborated by the numerical calculations which will be discussed in detail in Subsection~C of the article. 

Though the networks are characterized by the time reversal symmetry their missing level spectral statistics and the enhancement factor obey closer the Gaussian unitary ensemble predictions than the GOE ones. We performed the numerical calculations for open quantum graphs simulating microwave networks with no internal absorption to investigate the doublets which were not experimentally resolved. We show that the doublet size distribution is close to the Poisson distribution. 

One should point out that the previous experimental studies, in which microwave simulators were applied, were devoted to chaotic networks with the level spacing statistics belonging either to GOE, GUE or GSE statistics. In this article the networks and graphs characterized by the structure of nearly degenerate doublets are for the first time experimentally and numerically studied.

\section{Unidirectional quantum graphs} 

In Ref. \cite{Akila2015} the undirected quantum graph $\Gamma$ composed of two unidirectional graphs $\Gamma_{+} $ and $\Gamma_{-} $ has been theoretically and numerically considered. 
In this realization of the unidirectional graphs the following structure of the vertex scattering matrices $\hat{\sigma}_i$ has been proposed 

\begin{equation}\label{1}
	\hat{\sigma}_i=\left[\begin{array}{c@{\;}c@{\;}}
		
		\hat{0} & \hat{U}_i\\
		\hat{U}_i^{\dag} & \hat{0} \\

	\end{array}\right] \mbox{with } \hat{U}_i \hat{U}_i^{\dag}= \hat{U}_i^{\dag}\hat{U}_i=1,
\end{equation}
where $\hat{0}$ is the zero matrix with all the entries equal zero and $ \hat{U}_i$ is a square unitary matrix.
Due to the off-diagonal structure of $\hat{\sigma}_i$ the transition from the graph $\Gamma_{+} $ to $\Gamma_{-} $ and vice versa is impossible and dynamics on $\Gamma_{+} $ and $\Gamma_{-} $ are completely decoupled. The splitting of $\Gamma $ into $\Gamma_{+} $ and $\Gamma_{-} $ is only possible if the vertices have even degree, e.g., in Ref. \cite{Akila2015} the vertices $\hat{\sigma}_i$ with the valency $v_i=4$ have been considered. 

The theoretical investigations of unidirectional quantum graphs \cite{Akila2015} dealt only with close nondissipative systems. However, the real experimental systems are open and characterized by internal absorption. In this work we use microwave networks simulating quantum graphs to investigate properties of coupled unidirectional graphs $\Gamma_{+} $ and $\Gamma_{-} $. The internal coupling of the unidirectional graphs was enforced by the $T$-junctions, vertices with the valency $v_T=3$, which were introduced to couple the microwave analyzer via the external leads with the investigated network. Details of the experimental setup will be given in the next section.

\section{Coupled unidirectional microwave networks}

The properties of the coupled unidirectional quantum graphs were investigated experimentally using microwave networks \cite{Hul2004}. The scheme of an undirected microwave network simulating an undirected quantum graph $\Gamma $ composed of two coupled unidirectional graphs $\Gamma_{+} $ and $\Gamma_{-} $ is shown in Fig.~1. 

The network is constructed of SMA microwave cables and microwave joints that act as edges and vertices of the simulated quantum graph. A microwave cable consists of outer and inner conductors of radius 
$r_1=0.15$ cm and $r_2=0.05$ cm, respectively. The separation between two conductors is filled with Teflon having 
the dielectric constant $\epsilon= 2.06$. Five microwave hybrid couplers (RF-Lambda RFHB02G08GPI), vertices $\hat{\sigma_i}$ with $v_i=4$, are used to obtain the undirected network $\Gamma$ 
with the coupled unidirectional networks $\Gamma_{+} $ and $\Gamma_{-} $, denoted in Fig.~1 by red and blue arrows, respectively. 

The $T$-junctions play a double role. They couple the network to the measuring system via HP 85133-616 and HP 85133-617 flexible microwave cables (leads $\cal L$$^{\infty}_{1}$ and $\cal L$$^{\infty}_{2}$) and additionally, because they are undirected, the unidirectional networks $\Gamma_{+} $ and $\Gamma_{-} $ with each other.

The properties of the $T$-junction with Neumann boundary conditions are described by its scattering matrix 

\begin{equation}\label{2}
	\hat{\sigma}_{T}= \frac{1}{3}\left[\begin{array}{c@{\;}c@{\;}c@{\;}}
		
		-1 & 2 & 2\\
		2 & -1 & 2\\
		2 & 2 & -1\\		
	\end{array}\right].
\end{equation}

The coupling between the unidirectional networks $\Gamma_{+} $ and $\Gamma_{-} $ is possible because of backscattering, represented by the diagonal elements in $\hat{\sigma}_{T}$ scattering matrix.

In the case of unidirectional vertices $\hat{\sigma}_i$ (couplers RF-Lambda RFHB02G08GPI) their scattering matrices in the operating frequency range $\nu\in [2,8] $ GHz are the following

\begin{equation}\label{3}
	\hat{\sigma}_{i}=\frac{1}{\sqrt{2}}\left[\begin{array}{c@{\;}c@{\;}c@{\;}c@{\;}}
		
		0 & 0 & 1 & 1\\
		0 & 0 & -1 & 1\\
		1 & -1 & 0 & 0\\
		1 & 1 & 0 & 0\\

	\end{array}\right].
\end{equation}

The diagonal elements of $ \hat{\sigma}_{i}$ matrices are zeros preventing from backscattering and therefore also from coupling of the unidirectional networks $\Gamma_{+} $ and $\Gamma_{-} $. The scattering matrix $ \hat{\sigma}_{i}$ has a form of the scattering matrix presented in Definition (\ref{1}) with the unitary matrix $\hat{U}_i$ defined as follows

\begin{equation}\label{4}
\hat{U}_i= \frac{1}{\sqrt{2}}\left[\begin{array}{c@{\;}c@{\;}}
		
		1 & 1\\
		-1 & 1 \\

	\end{array}\right]. 
\end{equation}

 In order to keep a one-to-one quantum-microwave vertex analogy the microwave vertex scattering matrices $\hat{\sigma}_v(\nu)$ and $\hat{\sigma}_v(\nu_0)$ at frequencies $\nu$ and $\nu_0$ for Neumann boundary conditions \cite{Exner1989} should be related by the equation \cite{Kostrykin1999,Kurasov2010,Berkolaiko2013,Exner2018,Lawniczak2020a}:

\begin{equation} \label{5}
	\hat{\sigma}_v(\nu) = \frac{(\nu +\nu_0) \hat{\sigma}_v(\nu_0) + (\nu-\nu_0)\hat I}{(\nu+\nu_0)\hat I + (\nu-\nu_0) \hat{\sigma}_v (\nu_0)}.
\end{equation}
Here, the matrix $\hat I$ denotes the identity matrix of the dimension of the vertex scattering matrices $\hat{\sigma}_v(\nu)$ and $\hat{\sigma}_v(\nu_0)$.

It can be easily checked that for the components of the microwave network presented in Fig.~1, namely microwave $T$-junctions and couplers, the scattering matrices $\hat{\sigma}_{T}$ and $\hat{\sigma}_{i}$ are unitary and Hermitian, fulfilling Eq.~(\ref{5}).

The lengths of edges of the quantum graph are equivalent to the optical lengths of the edges of the microwave network, i.e., $l_{opt} = \sqrt{\epsilon}l_{ph}$, where $l_{ph}$ is the physical length of a network edge. The total optical length $L_{tot}$ of the network was $7.955 \pm 0.012$ m. The optical lengths of the edges of the network are the following: $ l_1 = 0.649 \pm 0.001$ m, $l_2 = 0.788 \pm 0.001 $ m, $l_3 = 1.142 \pm 0.001$ m, $l_4 = 0.382 \pm 0.001$ m, $l_5 = 0.513 \pm 0.001$ m, $l_6 = 0.435 \pm 0.001$ m, $l_7 = 0.787 \pm 0.001$ m, $l_8 = 0.480 \pm 0.001$ m, $l_9 = 0.760 \pm 0.001$ m, $l_{10} = 0.897 \pm 0.001$ m, $l_{11} = 0.657 \pm 0.001$ m, $l_{12} = 0.465 \pm 0.001$ m.
  
In order to obtain an ensemble of coupled unidirectional networks $\Gamma_{+} $ and $\Gamma_{-} $ the lengths of two bonds of the undirected $\Gamma$ network were changed by using the phase shifters PS1 and PS2 in such a way that the total optical length $L_{tot}$ of the network was kept constant. Due to couplers' frequency characteristics, the experiment was performed within the frequency range $\nu\in [2,8] $ GHz. In this interval according to the Weyl's law in each spectrum one should expect $\sim$318 resonances, however, due to the doublet structure of very closely degenerate resonances induced by the coupled unidirectional networks $\Gamma_{+} $ and $\Gamma_{-} $ in the experiment we observed at most half of them. In practice, in each of applied 50 network realizations about 4\% of resonances (not resolved doublets) were not detected.

In Fig.~2 we show a photograph of the experimental setup. It consists the microwave undirected network $\Gamma$ connected via HP 85133-616 and HP 85133-617 flexible microwave cables to a vector network analyzer (VNA), Agilent E8364B, to measure the two-port scattering matrix $\hat{S}(\nu)$ of the network.  The inset shows an example of the modulus of the diagonal scattering matrix element $|S_{11}(\nu)|$ of the network measured in the frequency range $5.24-5.74$ GHz. Despite of the coupled unidirectional networks $\Gamma_{+} $ and $\Gamma_{-} $ experimental resonances (local minima in $|S_{11}(\nu)|$, marked by vertical lines) remain doubly degenerate within the experimental resolution.

\subsection{Spectral statistics of the undirected microwave network $\Gamma$}

The spectral properties of the undirected microwave network $\Gamma$ were investigated using the most common measures of the short- and long-range spectral correlations: the nearest-neighbor spacing distribution $P(s)$ and the spectral rigidity $\Delta_3(L)$. In order to perform these analyses the resonance frequencies $\nu_i$ of the network were rescaled (unfolded) to eliminate system specific properties. Since experimental resonances are doubly degenerate this can be done using the Weyl's formula for the network with the total optical length $L_{tot}'=L_{tot}/2$. Then, the unfolded eigenvalues determined from the resonance frequencies $\nu_i$ are given by $\epsilon_i=L_{tot}\nu_i/c$, where $c$ is the speed of light in the vacuum.

The nearest-neighbor spacing distribution (NNSD) $P(s)$ describes the distribution of the spacings between adjacent eigenvalues $s_i=\epsilon_{i+1}-\epsilon_i$ in terms of their mean value $\langle s\rangle$, while the spectral rigidity $\Delta_3(L)$ corresponds to the least square deviation of the integrated spectral density of the unfolded $\epsilon_i$ from the straight line best fitting it in an interval of length $L$ \cite{Mehta1990}.

 The nearest-neighbor spacing distribution $P(s)$ which takes into account the incompleteness of a level sequence (missing levels) is given by~\cite{Bohigas2004}
\begin{equation} \label{6}
	P(s)= \sum_{n=0}^{\infty}(1-\phi)^{n}p(n,\frac{s}{\phi}).
\end{equation}
For complete sequences, $\phi$ = 1, $P(s) = p(0,s)$, which for GUE systems is well approximated by the Wigner surmise:
\begin{equation} \label{7}
	P(s) = \frac{32}{\pi^{2}}s^{2}\exp(-\frac{4}{\pi}s^{2}).
\end{equation}
For the fraction of observed levels $\phi< 1$ the following expression was used
$P(s) \simeq p(\frac{s}{\phi})+(1-\phi)p(1,\frac{s}{\phi}) + (1-\phi)^{2}p(2,\frac{s}{\phi})$, where \cite{Stoffregen1995}
\begin{equation} \label{8}
	p(n,\frac{s}{\phi})= \gamma (\frac{s}{\phi})^{\mu}\exp(-\kappa (\frac{s}{\phi})^{2}),
\end{equation}
with $\mu =7, 14$ for n= 1, 2, respectively, and $\gamma$ and $\kappa$ determined from the normalization conditions:
\begin{equation} \label{9}
	\int p(n,\frac{s}{\phi}) ds = \phi, \,\,\, \int sp(n,\frac{s}{\phi}) ds = \phi^2 (n+1).
\end{equation}

The spectral rigidity $\delta_3(L)$ in the case of $\phi< 1$ \cite{Bohigas2004} is given by 

\begin{equation} \label{10}
	\ \delta_3(L)= (1-\phi)\frac{L}{15}+ \phi^{2}\Delta_3(\frac{L}{\phi}),
\end{equation}

where for $\phi$ = 1 the spectral rigidity $\Delta_3(L)$ is defined by
\begin{equation} \label{11}
	\ \Delta_3(L)= \frac{L}{15} - \frac{1}{15L^{4}}\int_{0}^{L}(L-x)^{3}(2L^{2}-9xL-3x^{2})Y_{2}(x)dx.
\end{equation}
For GUE systems the two-point cluster function $Y_{2}(x)=(\frac{\sin \pi x}{\pi x})^{2}$ \cite{Mehta1990}.

The results for the discussed spectral measures are presented in Fig.~3. The NNSD and the spectral rigidity $\Delta_3(L)$ for the undirected microwave network $\Gamma$ is displayed in the panels (a) and (b), respectively. In Fig.~3(a) the experimental NNSD obtained using 7488 level spacings is presented by the green histogram. The experimental results are compared with the theoretical ones based on random matrix theory (RMT) for complete series of resonances $\phi = 1$ (GOE - black solid line, GUE - blue solid line) and the incomplete GUE one, with the fraction of observed levels $\phi=0.96$, 4\% of missing resonances, (red broken line), respectively. Fig.~3(a) shows that the experimental NNSD is shifted towards larger parameter $s$ in relation to the GUE distributions. The numerical analysis of a single unidirectional graph presented in Ref. \cite{Che2022} showed that the departure of its spectral characteristics from the GUE predictions was caused by not sufficiently complex wave dynamics in this graph. Therefore, also in our case the observed spectral deviation maybe attributed to not sufficiently complex wave dynamics in the coupled unidirectional graphs $\Gamma_{+} $ and $\Gamma_{-} $.  In Fig.~3(b) the spectral rigidity $\Delta_3(L)$ for the microwave network $\Gamma$ is presented by green circles. The experimental results are compared with the theoretical ones based on RMT for complete series of resonances $\phi = 1$, GOE - black solid line, GUE - blue solid line, and the incomplete series $\phi=0.96$ for GUE, red broken line, respectively. The inspection of the results reveals that the experimental results are close to the GUE missing level statistics with the fraction  $\phi=0.96$ of observed levels. 
 
\subsection{The elastic enhacement factor of the microwave network $\Gamma$}

The measurement of the two-port scattering matix $\hat{S}$ of the undirected network $\Gamma $ allows for the evaluation of 
the elastic enhancement
factor \cite{Fyodorov2004,Savin2005}
\begin{equation}
	\label{12}
	W_S=\frac{\sqrt{\mbox{var}(S_{11})\mbox{var}(S_{22})}}{\mbox{var}(S_{12})},
\end{equation}
where, e.g., $\mbox{var}(S_{12}) \equiv \langle |S_{12}|^2\rangle
-|\langle S_{12} \rangle |^2$ stands for the variance of the
matrix element $S_{12}$.
The diagonal elements of the scattering matrix $\hat S$ can be parameterized as
	 $S_{ii}=\sqrt{R_i}e^{i\theta_i}$,
where $R_i$ and $\theta_i$ are the reflection coefficient and the phase measured at the $i^{th}$ port of the network.

The elastic enhancement
factor $W_S$ is parametrized by the dimensionless parameter $\gamma=2\pi \Gamma_W /\Delta_S $, characterizing the absorption strength \cite{Fyodorov2004,Savin2005}, where
$\Gamma_W$ and $\Delta_S$ are the width of resonances and the mean level
spacing, respectively. It is important to point out that system characteristics defined by the scattering matrix are not sensitive on missing levels.
 It was established for GOE ($\beta=1$) and GUE ($\beta=2$) systems that the elastic enhancement factor for weak absorption $\gamma \ll 1$ approaches the limit of $W_S=2/\beta +1$, while in the case of strong absorption $\gamma \gg 1$ the limit is $W_S=2/\beta$.

Because the properties of the elastic enhancement factor $W_S$ strongly depend on $\mathcal{T}$-symmetry of the system it can be used as a sensitive measure of time invariance violation. 

In such a situation the effective parameter $\gamma$ can be evaluated using the distribution $P(R)$ of the reflection coefficient $R$.
For systems without $\mathcal{T}$-invariance ($\beta=2$), the
analytic expression for the distribution of the reflection
coefficient $R$ is given by \cite{Beenakker2001,Savin2005}

\begin{equation}
	\label{13}
	P(R)=\frac{2}{(1-R)^2}P_0\Bigl(\frac{1+R}{1-R}\Bigr),
\end{equation}
where $P_0(x)$ is the probability distribution defined by

\begin{equation}
	\label{14}
	P_0(x) =\frac{1}{2}\Bigl[A\Bigl(\frac{\alpha(x+1)}{2}\Bigr)^{\beta/2} + B\Bigr]\exp\Bigl(-\frac{\alpha(x+1)}{2}\Bigr),
\end{equation}
where $\alpha=\gamma \beta/2$, $A=e^{\alpha}-1$ and $B=1+\alpha-e^{\alpha}$.

The probability distribution $P_0(x)$ can be also applied for calculating the distribution of the imaginary part $P(v)$ 
of the diagonal elements of the Wigner's $\hat K$ matrix \cite{Fyodorov2004}
\begin{equation}
	\label{15} P(v)=\frac{\sqrt{2}}{\pi
		v^{3/2}}\int^{\infty}_{0}dqP_0\Bigl[q^2+\frac{1}{2}\Bigl(v+\frac{1}{v}\Bigr)\Bigr].
\end{equation}
 The
 distribution $P(v)$ is known in solid-state physics as the local density of states (LDoS) \cite{Fyodorov2004}.

For each realization of the network $\Gamma$ the absorption strength $\gamma=\frac{1}{2}\sum_{i=1}^2\gamma_{i}$ was experimentally
evaluated by adjusting the theoretical mean reflection coefficient
\begin{equation}
	\label{16} \langle R\rangle ^{th} = \int _0^1dRRP(R),
\end{equation}
to the experimental one $\langle R_{i} \rangle$
 obtained after eliminating the direct
processes \cite{Fyodorov2005,Kuhl2005,Hul2005}. Here the index $i=1,2$ denotes the
port $1$ or $2$.

 In Fig.~4 we show the experimental distributions $P(R)$ of the reflection
coefficient $R$ for the microwave network $\Gamma$ at three values of the absorption strength $\gamma =3.6 \pm 0.6$, $4.6 \pm 0.3$, and $6.4 \pm 0.3$. They are marked by black, red, and green open circles, respectively. The measurements were done in the frequency ranges $\nu \in [2,4]$, $ [4,6]$, and $[6,8]$ GHz, respectively, and were averaged over 500 microwave network realizations. The values of the absorption strength $\gamma$ were assigned to the experimental curves by fitting the theoretical distributions $P(R)$ 
			calculated from Eq.~(\ref{13}) with the absorption coefficients
			$\gamma =3.6$, $4.6$, and $6.4$, marked by black, red, and green solid lines,
			respectively.	 

Fig.~5 shows the elastic enhancement factor $W_S$ (black open circles) evaluated experimentally for the undirected mirowave network $\Gamma$ in the frequency range $\nu \in [2,8]$ GHz. The experimental results were averaged in 1~GHz window over 300 microwave network realizations. In this frequency range the averaged absorption strength parameter $\gamma=4.9\pm 0.4$. 
		The experimental results are compared to the expected theoretical values of $W_S$ for GUE systems which are marked by red solid line. The experimental elastic enhancement factor is on average slightly higher that the theoretical one predicted for GUE systems. Also this discrepancy is probably caused by not sufficiently complex wave dynamics in the microwave network $\Gamma$. The black broken lines in Fig.~5 show the lowest $ W_S=1$ and the highest $W_S=2$ theoretical limits 	of the elastic enhancement factor predicted for GUE systems.
		
In Fig.~6 we show the experimental distribution $P(v)$ of the imaginary part 
			of the diagonal elements of the Wigner's $\hat K$
			matrix for the microwave undirected network $\Gamma$ at $\gamma =4.9$ (black open circles), averaged over 500 microwave network realizations, for the frequency range $\nu \in [2,8]$ GHz. The experimental results are compared with the theoretical
			distribution $P(v)$ evaluated from Eq.~(\ref{15}) for $\gamma =4.9$ (red solid line). The agreement between the experimental and theoretical results is good. 
	
\subsection{Numerical analysis of doublets properties}

To investigate numerically the properties of doublets, which were not experimentally resolved, the microwave undirected network $\Gamma$ was simulated in the calculations by the open, dissipationless quantum graph $\Gamma$. 
The secular function $\xi$, whose zeros define the spectrum of the
graph was expressed by using the method of pseudo-orbits \cite{Band2012,Lipovsky2015,Lipovsky2016,Lawniczak2019}

\begin{equation}\label{17}
\xi=	\mbox{det}\left[ \hat{I}_{2N}-\hat{L}\hat{S}_{G}\right], 
\end{equation} 
where $\hat{I}_{2N}$ is $2N \times 2N$ identity matrix, $N$ is the number of the internal edges of the graph and $\hat{L}=\mbox{diag}\left[\exp(ikl_{1}),...,\exp(ikl_{N}),\exp(ikl_{1}),...,\exp(ikl_{N})\right]$, $l_{1}...l_{N}$ are the lengths of the respective edges of the graph. The $\hat{S}_{G}$ matrix, called the bond-scattering matrix \cite{Band2012}, contains scattering conditions at the graph vertices.
The full form of the $\hat{S}_{G}$ matrix is specified in the Appendix.

The nearest-neighbor spacing distribution $P(s)$ and the spectral rigidity $\Delta_3(L)$ of the graph $\Gamma$ consisting of the coupled unidirectional graphs $\Gamma_{+} $ and $\Gamma_{-} $ are shown in Fig.~7(a) and Fig.~7(b), respectively. In this case the unfolded eigenvalues were determined from the resonance frequencies $\nu_i$ applying the Weyl's formula $\epsilon_i=2L_{tot}\nu_i/c$, where $L_{tot}$ is the total optical length of the graph. The numerical calculations were performed in the frequency range $\nu=[2,8]$ GHz and were averaged over 50 configurations of the graph $\Gamma$. 

Because of the doublet structure of the spectra the nearest-neighbor spacing distribution $P(s)$, prepared using 15850 level spacings (green histogram), displays a large peak at small values of the parameter $s$ and is significantly different from the Poisson (red dotted-dashed line), GOE (black dotted line), and GUE (blue dashed line) distributions, respectively. 

 The inset in Fig.~7(a) shows an example of the spectrum of the undirected graph $\Gamma$ (coupled unidirectional graphs $\Gamma_+$ and $\Gamma_-$) calculated in the frequency range $3.0-3.3$ GHz. Because the size of the doublets are very small, between $0.3 - 3.9$ MHz, they are not resolved within the experimental resolution (15 MHz) and are not observed experimentally.

The spectral rigidity $\Delta_3(L)$ of the graph $\Gamma$ (green circles) presented in Fig.~7(b) due to the presence of doublets increases faster for small $L$ than the Poisson one (red dotted-dashed line) and only for $L>8$ slowly saturates at the value $\Delta_3(L) \approx 0.5$ which is significantly higher than the GUE (blue dashed line) and GOE (black dotted line) predictions for the presented range of $L \le 20$.  

In Fig.~8(a) and Fig.~8(b) we show the nearest-neighbor spacing distribution $P(s)$ and the spectral rigidity $\Delta_3(L)$ of the graph $\Gamma$ obtained under the assumption that the doublets are not resolved and are treated as singlet states. In this case, similarly to the experimental situation, we assumed that the unfolded resonances were determined from the Weyl's formula $\epsilon_i=L_{tot}\nu_i/c$. 

In Fig.~8(a) the numerical distribution $P(s)$ obtained for the graph $\Gamma$ (green histogram) is compared with the distribution $P(s)$ evaluated for the simplified, closed graph $\Gamma'$, composed of 10 edges and 5 couplers (red histogram). Both distributions were made using 7900 level spacings. The graph $\Gamma'$ was obtained from the graph $\Gamma $, presented in Fig.~1, by removing of two $T$-junctions. The quantum graph $\Gamma'$ is characterized by the spectrum of exactly doubly degenerate states. The two mentioned above distributions $P(s)$ are compared with the Poisson (red dotted-dashed line), GOE (black dotted line), and GUE (blue dashed line) distributions, respectively. Both numerical distibutions $P(s)$ are close to the GUE distribution, however, they are slightly shifted towards larger values of the level spacing $s$. Moreover, the distribution $P(s)$ of the simplified graph $\Gamma'$ is more localized around the center of the GUE distribution than the one for the graph $\Gamma$, suggesting that backscattering present at $T$-junctions of the graph $\Gamma$ causes some additional deviations from the GUE distribution. 	 

The spectral rigidity $\Delta_3(L)$ presented in Fig.~8(b) shows a significant deviation from the GUE prediction.  Our results corroborate the observation reported in Ref. \cite{Che2022} that the unidirectional graphs may not generate a wave dynamics of sufficient complexity to accurately reproduce RMT predictions.  

In Fig.~9 we show the doublet size distribution $P(\Delta)$ (blue histogram) of the graph $\Gamma$ consisting of the coupled unidirectional graphs $\Gamma_{+} $ and $\Gamma_{-} $ (see Fig.~1). The doublet size was normalized to the mean value $\langle\Delta \rangle =1$. In the calculation of the distribution  $P(\Delta)$  7900 doublets were used.
The distribution $P(\Delta)$ is compared to the Poisson distribution $P_{Poisson} = \exp( -\Delta)$ (red solid line). Fig.~9 demonstrates that the distribution $P(\Delta)$ is close to the Poisson one.

\section {Summary and conclusions}

We investigated experimentally undirected open microwave network $\Gamma $ with internal absorption composed of two coupled unidirectional networks $\Gamma_{+} $ and $\Gamma_{-} $ corresponding to two possible directions of motion on their edges. The two-port scattering matrices of the network were measured. The comparison of the number of experimental resonances with the theoretical one predicted by the Weyl's law showed that the resonances are doubly degenerate. Though the networks are characterized by the time reversal symmetry their missing level nearest-neighbor spacing distribution $P(s)$ and the spectral rigidity $\Delta_3(L)$ ($\phi=0.96$) do not obey the GOE predictions. The missing level NNSD $P(s)$ reminds shifted towards larger values of mean level spacing $s$  GUE distribution while the missing level spectral rigidity $\Delta_3(L)$ is in good agreement with the missing level prediction for GUE. Furthermore, the distributions of the reflection coefficient and the imaginary part of the Wigner's reaction matrix as well as the enhancement factor of the networks were evaluated. The aforementioned characteristics of chaotic systems are defined by the scattering matrix of the network. Therefore, they are not sensitive on missing levels. The obtained results are close to the GUE prediction, though the experimental enhancement factor appears to be slightly above it. We used the numerical calculations for open quantum graphs simulating microwave networks with no internal absorption to investigate their spectral statistics and doublets which were not experimentally resolved. The numerically obtained spectral characteristics show significant deviations from the GUE predictions. We show that the doublet size distribution is close to the Poisson distribution. Reported in this paper discrepancies between the experimental results and the GUE ones as well as between the numerical results simulating the experimental ones and the GUE predictions maybe associated with the presence of backscattering and not sufficiently complex wave dynamics on the coupled unidirectional networks and graphs $\Gamma_{+} $ and $\Gamma_{-} $.

\section{Acknowledgments}

This work was supported in part by the National Science Centre, Poland, Grant No. 2018/30/Q/ST2/00324. We are grateful to S. Bauch for critical reading of the manuscript.

\section{Appendix}

\centerline{\bf Numerical calculations}

The resonance conditions for the unidirectional graphs can be expressed using the method of pseudo-orbits \cite{Band2012,Lipovsky2015,Lipovsky2016,Lawniczak2019}, by solving the equation:
\begin{equation}\label{18}
\mbox{det}\left[ \hat{I}_{2N}-\hat{L}\hat{S}_{G}\right]=0,
\end{equation}

where $\hat{I}_{2N}$ is $2N$x$2N$ identity matrix, $N$ is the number of the internal edges of the graph $\left(N=12\right)$ and $\hat{L}=diag\left[exp(ikl_{1}),...,exp(ikl_{12}),exp(ikl_{1}),...,exp(ikl_{12})\right]$, $l_{1}...l_{12}$ are the lengths of the respective arms, according to figure 1 in the paper. The $\hat{S}_{G}$ matrix, called the bond-scattering matrix \cite{Band2012}, contains scattering conditions at the network vertices (see Fig.~1), i.e., five vertices with the valency $v_T=4$ and the Neumann boundary conditions, ensuring that the graph $\Gamma$ contains two unidirectional graphs $\Gamma_+$ and $\Gamma_-$,

\begin{equation}\label{19}
	\hat{\sigma}_{i}= \frac{1}{\sqrt{2}} \left[\begin{array}{c@{\;}c@{\;}c@{\;}c@{\;}}
		
		0 & 0 & 1 & 1\\
		0 & 0 & -1 & 1\\
		1 & -1 & 0 & 0\\
		1 & 1 & 0 & 0

	\end{array}\right],
\end{equation}

\smallskip

and two T-junction vertices with the valency $v_T=3$ and with the Neumann boundary conditions:

\begin{equation}\label{20}
	\hat{\sigma}_{T}= \frac{1}{3} \left[\begin{array}{c@{\;}c@{\;}c@{\;}}
		
		-1 & 2 & 2\\
		2 & -1 & 2\\
		2 & 2 & -1\\		
	\end{array}\right].
\end{equation}

\smallskip

The backscattering introduced by the T-junctions (presence of the diagonal elements in 	$\hat{\sigma}_{T}$) causes that the resonances of the graphs $\Gamma_+$ and $\Gamma_-$ are not doubly degenerate. 

The matrix $\hat{S}_{G}$ has a form:

\fontsize{10}{10}
\begin{equation}\label{4}
	\hat{S}_{G}=\left[\begin{array}{c@{\;}c@{\;}c@{\;}c@{\;}c@{\;}c@{\;}c@{\;}c@{\;}c@{\;}c@{\;}c@{\;}c@{\;}c@{\;}c@{\;}c@{\;}c@{\;}c@{\;}c@{\;}c@{\;}c@{\;}c@{\;}c@{\;}c@{\;}c@{\;}}
		
		0 & 0 & 0 & 0 & 0 & 0 & 0 & 0 & 0 & 0 & \frac{2}{3} & 0 & \frac{-1}{3} & 0 & 0 & 0 & 0 & 0 & 0 & 0 & 0 & 0 & 0 & 0\\
		\frac{1}{\sqrt{2}} & 0 & 0 & 0 & 0 & 0 & 0 & \frac{1}{\sqrt{2}} & 0 & 0 & 0 & 0 & 0 & 0 & 0 & 0 & 0 & 0 & 0 & 0 & 0 & 0 & 0 & 0\\
		0 & \frac{1}{\sqrt{2}} & 0 & 0 & 0 & 0 & 0 & 0 & \frac{1}{\sqrt{2}} & 0 & 0 & 0 & 0 & 0 & 0 & 0 & 0 & 0 & 0 & 0 & 0 & 0 & 0 & 0\\
		0 & 0 & 0 & 0 & 0 & 0 & 0 & 0 & 0 & 0 & 0 & \frac{2}{3} & 0 & 0 & 0 & \frac{-1}{3} & 0 & 0 & 0 & 0 & 0 & 0 & 0 & 0\\
		0 & 0 & 0 & 0 & 0 & 0 & \frac{1}{\sqrt{2}} & 0 & 0 & \frac{1}{\sqrt{2}} & 0 & 0 & 0 & 0 & 0 & 0 & 0 & 0 & 0 & 0 & 0 & 0 & 0 & 0\\
		0 & 0 & \frac{-1}{\sqrt{2}} & \frac{1}{\sqrt{2}} & 0 & 0 & 0 & 0 & 0 & 0 & 0 & 0 & 0 & 0 & 0 & 0 & 0 & 0 & 0 & 0 & 0 & 0 & 0 & 0\\
		\frac{1}{\sqrt{2}} & 0 & 0 & 0 & 0 & 0 & 0 & \frac{-1}{\sqrt{2}} & 0 & 0 & 0 & 0 & 0 & 0 & 0 & 0 & 0 & 0 & 0 & 0 & 0 & 0 & 0 & 0\\
		0 & 0 & \frac{1}{\sqrt{2}} & \frac{1}{\sqrt{2}} & 0 & 0 & 0 & 0 & 0 & 0 & 0 & 0 & 0 & 0 & 0 & 0 & 0 & 0 & 0 & 0 & 0 & 0 & 0 & 0\\
		0 & 0 & 0 & 0 & \frac{1}{\sqrt{2}} & \frac{1}{\sqrt{2}} & 0 & 0 & 0 & 0 & 0 & 0 & 0 & 0 & 0 & 0 & 0 & 0 & 0 & 0 & 0 & 0 & 0 & 0\\
		0 & \frac{-1}{\sqrt{2}} & 0 & 0 & 0 & 0 & 0 & 0 & \frac{1}{\sqrt{2}} & 0 & 0 & 0 & 0 & 0 & 0 & 0 & 0 & 0 & 0 & 0 & 0 & 0 & 0 & 0\\
		0 & 0 & 0 & 0 & \frac{1}{\sqrt{2}} & \frac{-1}{\sqrt{2}} & 0 & 0 & 0 & 0 & 0 & 0 & 0 & 0 & 0 & 0 & 0 & 0 & 0 & 0 & 0 & 0 & 0 & 0\\
		0 & 0 & 0 & 0 & 0 & 0 & \frac{1}{\sqrt{2}} & 0 & 0 & \frac{1}{\sqrt{2}} & 0 & 0 & 0 & 0 & 0 & 0 & 0 & 0 & 0 & 0 & 0 & 0 & 0 & 0\\
		0 & 0 & 0 & 0 & 0 & 0 & 0 & 0 & 0 & 0 & 0 & 0 & 0 & \frac{1}{\sqrt{2}} & 0 & 0 & 0 & 0 & \frac{1}{\sqrt{2}} & 0 & 0 & 0 & 0 & 0\\
		0 & 0 & 0 & 0 & 0 & 0 & 0 & 0 & 0 & 0 & 0 & 0 & 0 & 0 & \frac{1}{\sqrt{2}} & 0 & 0 & 0 & 0 & 0 & 0 & \frac{-1}{\sqrt{2}} & 0 & 0\\
		0 & 0 & 0 & 0 & 0 & 0 & 0 & 0 & 0 & 0 & 0 & 0 & 0 & 0 & 0 & 0 & 0 & \frac{-1}{\sqrt{2}} & 0 & \frac{1}{\sqrt{2}} & 0 & 0 & 0 & 0\\
		0 & 0 & 0 & 0 & 0 & 0 & 0 & 0 & 0 & 0 & 0 & 0 & 0 & 0 & 0 & 0 & 0 & \frac{1}{\sqrt{2}} & 0 & \frac{1}{\sqrt{2}} & 0 & 0 & 0 & 0\\
		0 & 0 & 0 & 0 & 0 & 0 & 0 & 0 & 0 & 0 & 0 & 0 & 0 & 0 & 0 & 0 & 0 & 0 & 0 & 0 & \frac{1}{\sqrt{2}} & 0 & \frac{1}{\sqrt{2}} & 0\\
		0 & 0 & 0 & 0 & 0 & 0 & 0 & 0 & 0 & 0 & 0 & 0 & 0 & 0 & 0 & 0 & 0 & 0 & 0 & 0 & \frac{1}{\sqrt{2}} & 0 & \frac{-1}{\sqrt{2}} & 0\\
		0 & 0 & 0 & 0 & 0 & 0 & 0 & 0 & 0 & 0 & 0 & 0 & 0 & 0 & 0 & 0 & \frac{1}{\sqrt{2}} & 0 & 0 & 0 & 0 & 0 & 0 & \frac{1}{\sqrt{2}}\\
		0 & 0 & 0 & 0 & 0 & 0 & 0 & 0 & 0 & 0 & 0 & 0 & 0 & \frac{1}{\sqrt{2}} & 0 & 0 & 0 & 0 & \frac{-1}{\sqrt{2}} & 0 & 0 & 0 & 0 & 0\\
		0 & 0 & 0 & 0 & 0 & 0 & 0 & 0 & 0 & 0 & 0 & 0 & 0 & 0 & \frac{1}{\sqrt{2}} & 0 & 0 & 0 & 0 & 0 & 0 & \frac{1}{\sqrt{2}} & 0 & 0\\
		0 & 0 & 0 & 0 & 0 & 0 & 0 & 0 & 0 & 0 & 0 & 0 & 0 & 0 & 0 & 0 & \frac{-1}{\sqrt{2}} & 0 & 0 & 0 & 0 & 0 & 0 & \frac{1}{\sqrt{2}}\\
		0 & 0 & 0 & 0 & 0 & 0 & 0 & 0 & 0 & 0 & \frac{-1}{3} & 0 & \frac{2}{3} & 0 & 0 & 0 & 0 & 0 & 0 & 0 & 0 & 0 & 0 & 0\\
		0 & 0 & 0 & 0 & 0 & 0 & 0 & 0 & 0 & 0 & 0 & \frac{-1}{3} & 0 & 0 & 0 & \frac{2}{3} & 0 & 0 & 0 & 0 & 0 & 0 & 0 & 0
		
	\end{array}\right]
\end{equation}

\section{References}
\bibliography{modifiedMyRef}

\pagebreak

%\centerline {\bf Figure Captions}

\smallskip

\begin{figure}[tb]
	\begin{center}
		\hspace*{-2cm} 
		\rotatebox{0}{\includegraphics[width=1.2\textwidth,
			height=1.2\textheight, keepaspectratio]{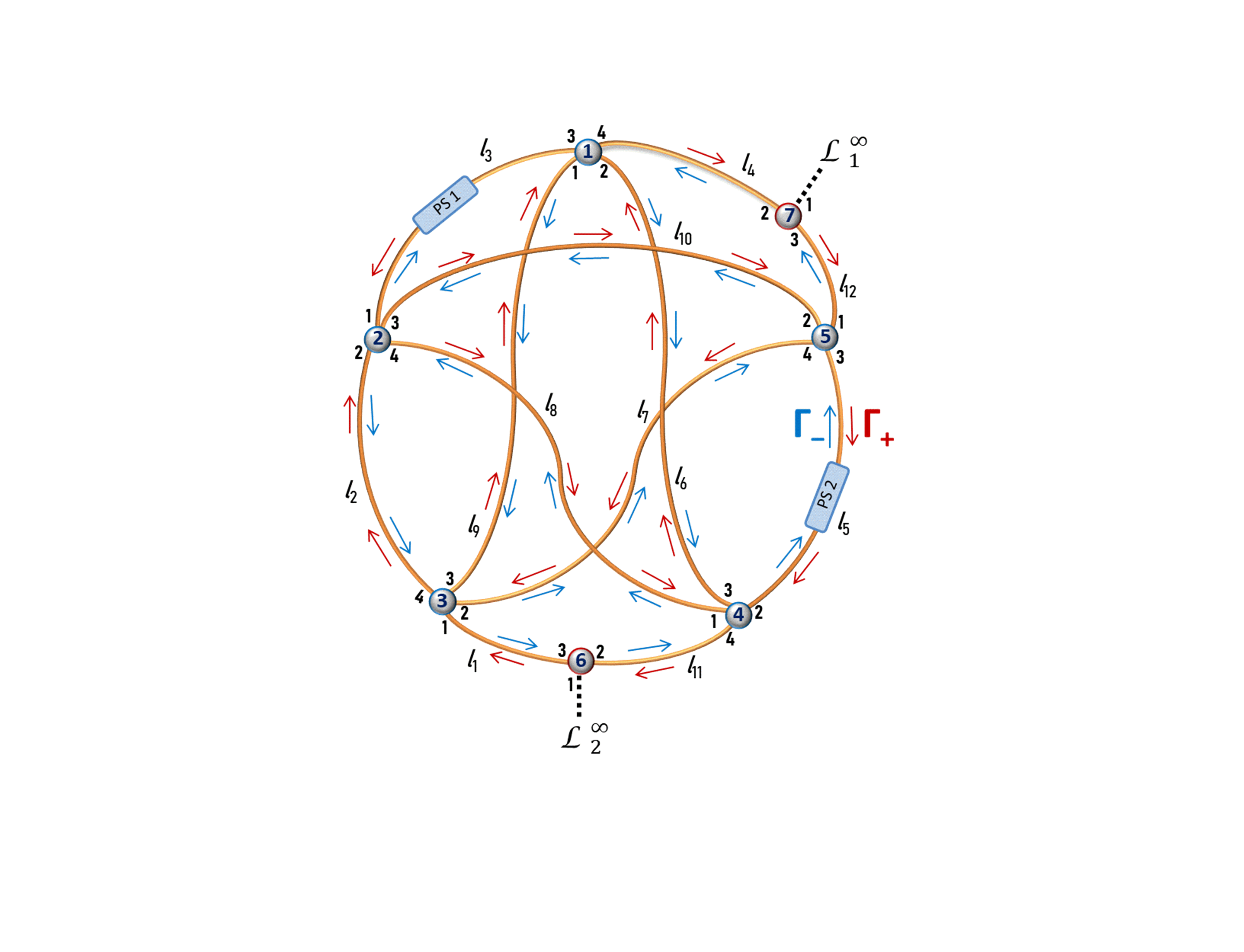}}
		\caption{The scheme of an undirected microwave network $\Gamma $ simulating an undirected quantum graph possessing two coupled unidirectional graphs $\Gamma_{+} $ (red arrows) and $\Gamma_{-} $ (blue arrows). The network contains two $T$-junctions (vertices 6 and 7), five microwave hybrid couples (vertices Nos. 1-5) and two phase shifters: PS1 and PS2. The network $\Gamma $ was connected at $T$-junctions via HP 85133-616 and HP 85133-617 flexible microwave cables (leads $\cal L$$^{\infty}_{1}$ and $\cal L$$^{\infty}_{2}$) to the ports of the vector network analyzer Agilent E8364B in order to measure the two-port scattering matrix $\hat{S}(\nu)$ of the network. 
		}\label{Fig1}
	\end{center}
\end{figure}

\begin{figure}[tb]
	\begin{center}
		\rotatebox{0}{\includegraphics[width=1.0\textwidth,
			height=1.0\textheight, keepaspectratio]{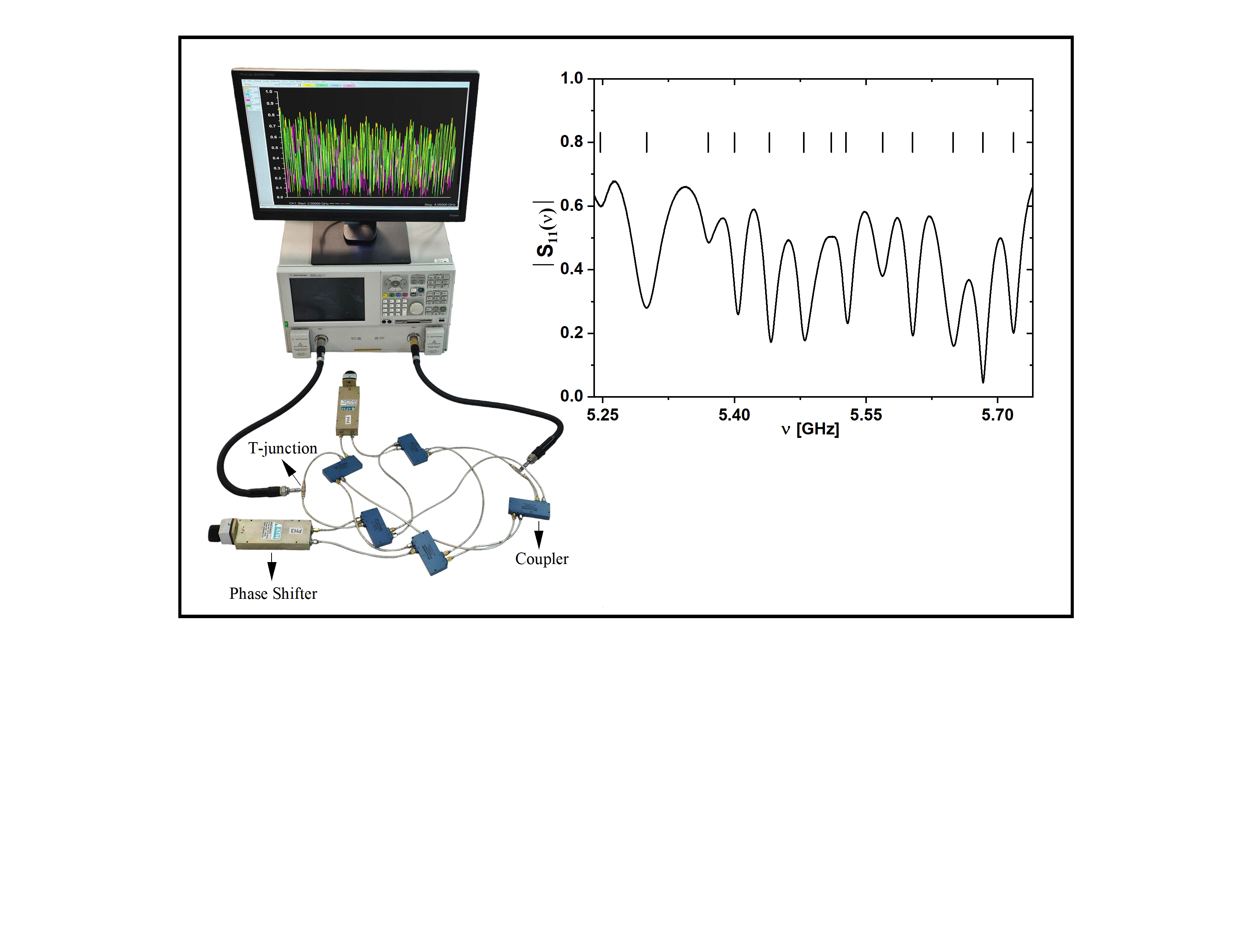}}
		\caption{A photograph of the experimental setup. It consists the microwave undirected network $\Gamma$ connected via HP 85133-616 and HP 85133-617 flexible microwave cables to a vector network analyzer (VNA), Agilent E8364B, to measure the two-port scattering matrix $\hat{S}(\nu)$ of the network.  The inset shows an example of the modulus of the diagonal scattering matrix element $|S_{11}(\nu)|$ of the network measured in the frequency range $5.24-5.74$ GHz. Due to the coupled unidirectional networks $\Gamma_{+} $ and $\Gamma_{-} $ every experimental resonance (marked by vertical lines) remains doubly degenerate within the experimental resolution.		 
		}\label{Fig2}
	\end{center}
\end{figure}

\begin{figure}[tb]
	\begin{center}
		\rotatebox{0}{\includegraphics[width=1.2\textwidth,
			height=1.2\textheight, keepaspectratio]{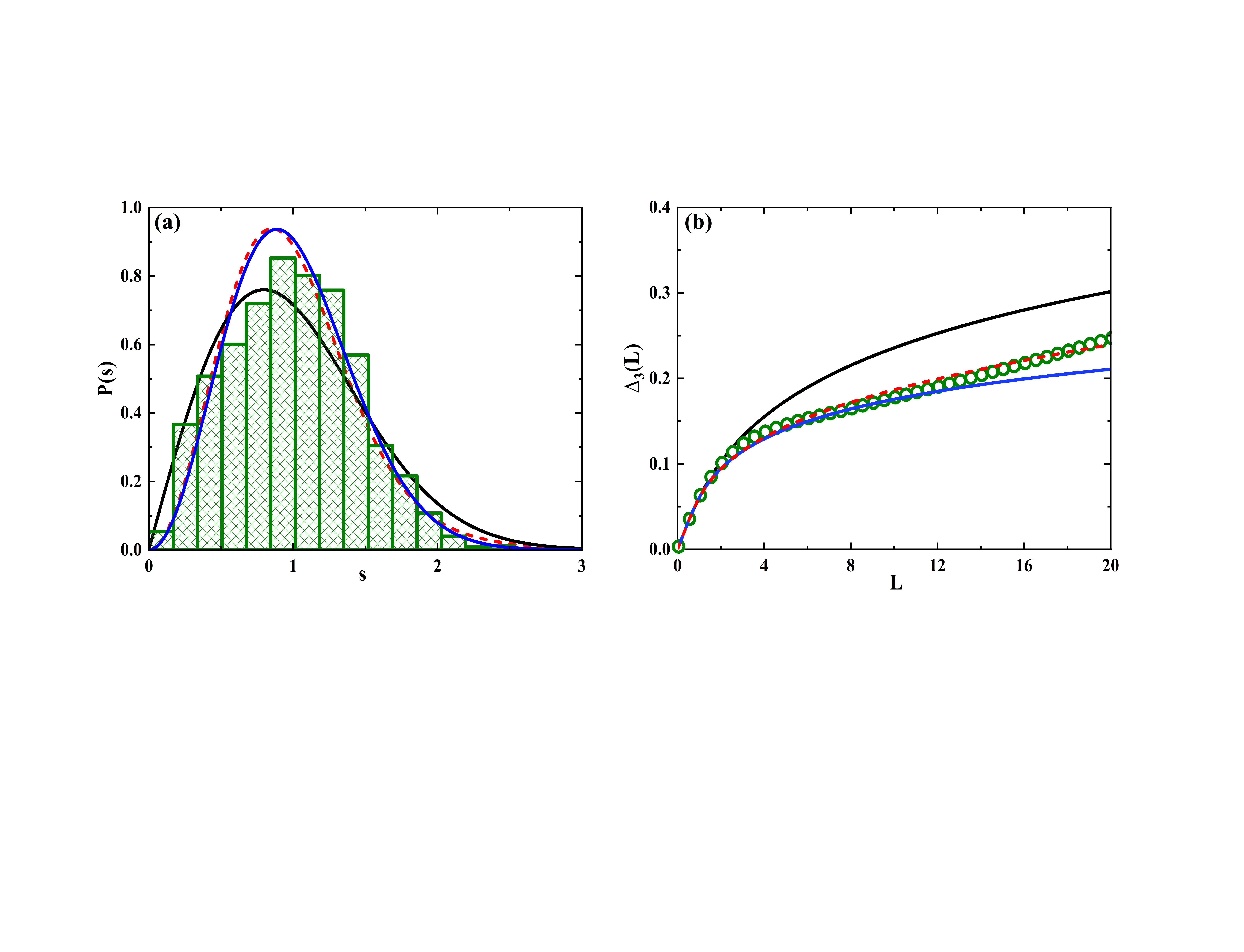}}
		\caption{(a) Experimental nearest-neighbor spacing distribution (green histogram) obtained for the undirected microwave networks $\Gamma$ (see Fig.~1). The experimental results are compared with the theoretical ones based on RMT for complete series of resonances $\phi = 1$ (GOE - black solid line, GUE - blue solid line) and the incomplete one $\phi=0.96$, 4\% of missing resonances, for GUE (red broken line), respectively. (b) The spectral rigidity for the microwave network $\Gamma$ is presented by green circles. The experimental results are compared with the theoretical ones based on RMT for complete series of resonances $\phi = 1$, GOE - black solid line, GUE - blue solid line, and the incomplete series $\phi=0.96$ for GUE, red broken line, respectively.		 
		}\label{Fig3}
	\end{center}
\end{figure}

\begin{figure}[tb]
\begin{center}
	\rotatebox{0}{\includegraphics[width=0.8\textwidth,
		height=0.8\textheight, keepaspectratio]{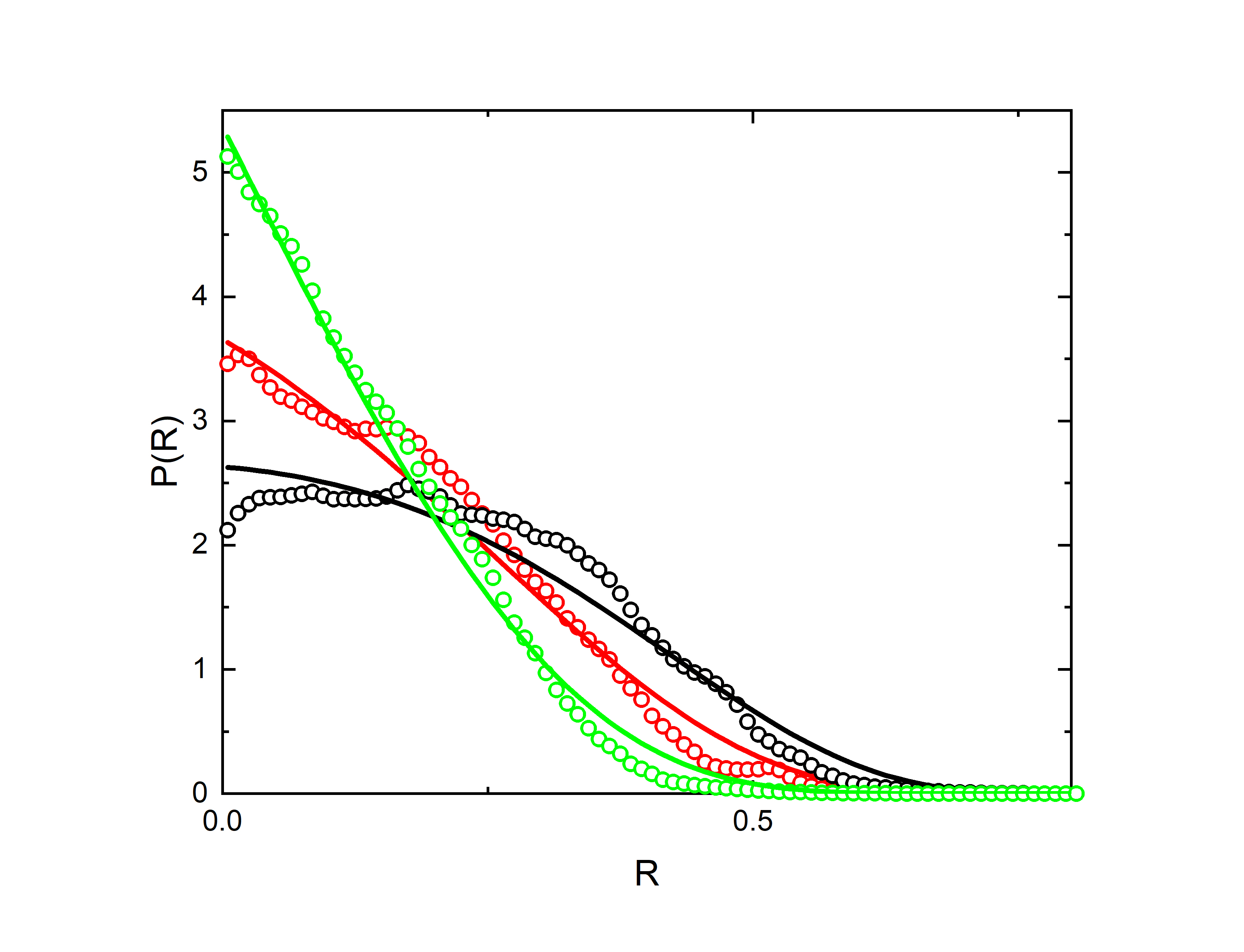}}
	\caption{Experimental distributions $P(R)$ of the reflection
		coefficient $R$ for the microwave network $\Gamma$ at $\gamma =3.6 \pm 0.6$ (black open circles, frequency range $\nu \in [2,4]$ GHz), $\gamma = 4.6 \pm 0.3$
		(red open circles, $\nu \in [4,6]$ GHz) and $\gamma=6.4 \pm 0.3$ (green open circles, $\nu \in [6,8]$ GHz) . The theoretical distributions $P(R)$
		calculated from the Eq.~(\ref{13}) are marked by black
		($\gamma =3.6$), red ($\gamma =4.6$), and green ($\gamma =6.4$) solid lines,
		respectively.	 
	}\label{Fig4}
\end{center}
\end{figure}

\begin{figure}[tb]
	\begin{center}
		\rotatebox{0}{\includegraphics[width=0.8\textwidth,
			height=0.8\textheight, keepaspectratio]{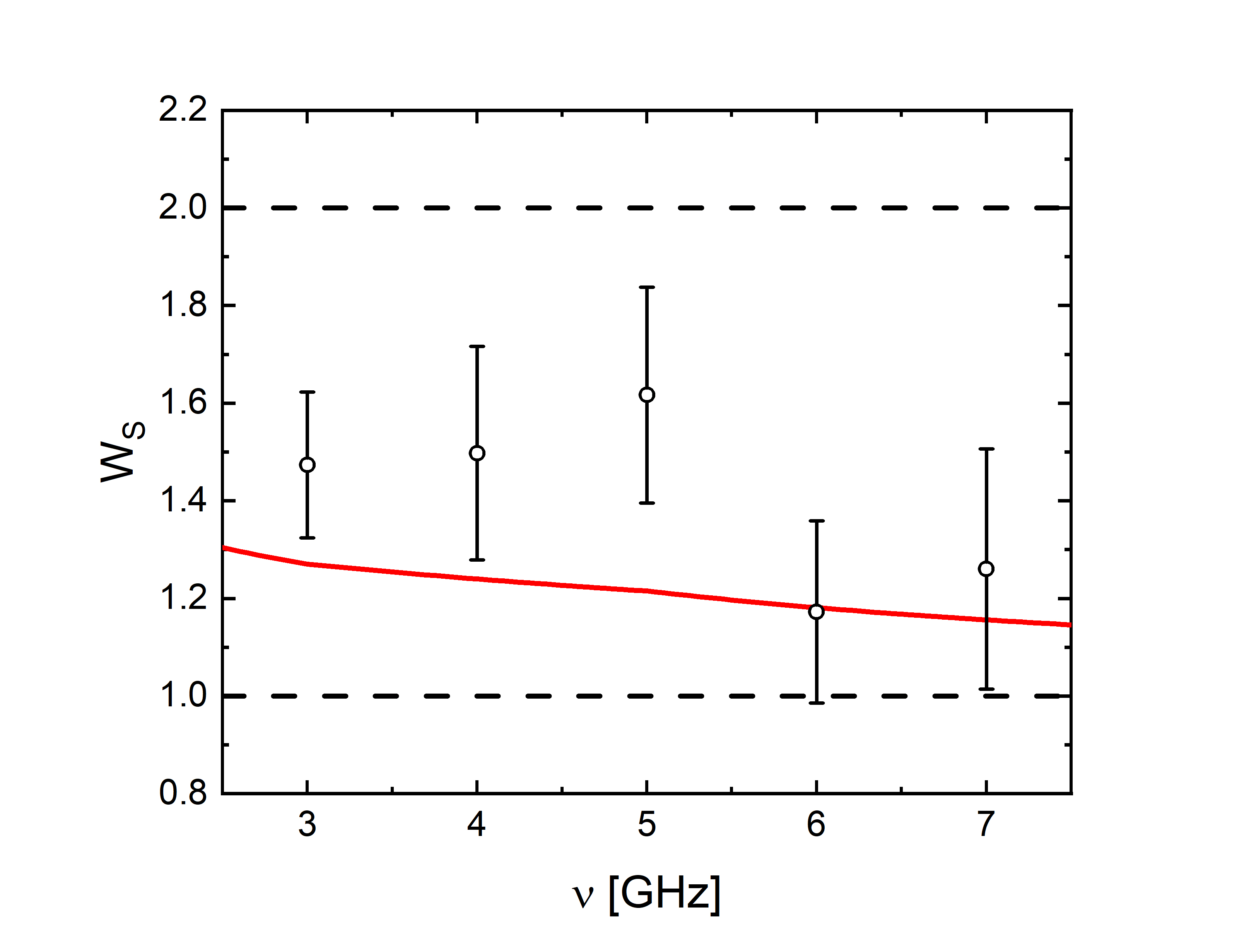}} 
		\caption{The elastic enhancement factor $W_S$ (black open circles) evaluated experimentally for the undirected mirowave network $\Gamma$ in the frequency range $\nu \in [2,8]$ GHz. In this frequency range the averaged absorption strength parameter $\gamma=4.9 \pm 0.4$.
			 The expected theoretical values for GUE systems are marked by red solid line.	The black broken lines show the lowest $ W_S=1$ and the highest $W_S=2$ theoretical limits 	of the elastic enhancement factor predicted for GUE systems.
		}\label{Fig5}
	\end{center}
\end{figure}

\begin{figure}[tb]
	\begin{center}
		\rotatebox{0}{\includegraphics[width=0.8\textwidth,
			height=0.8\textheight, keepaspectratio]{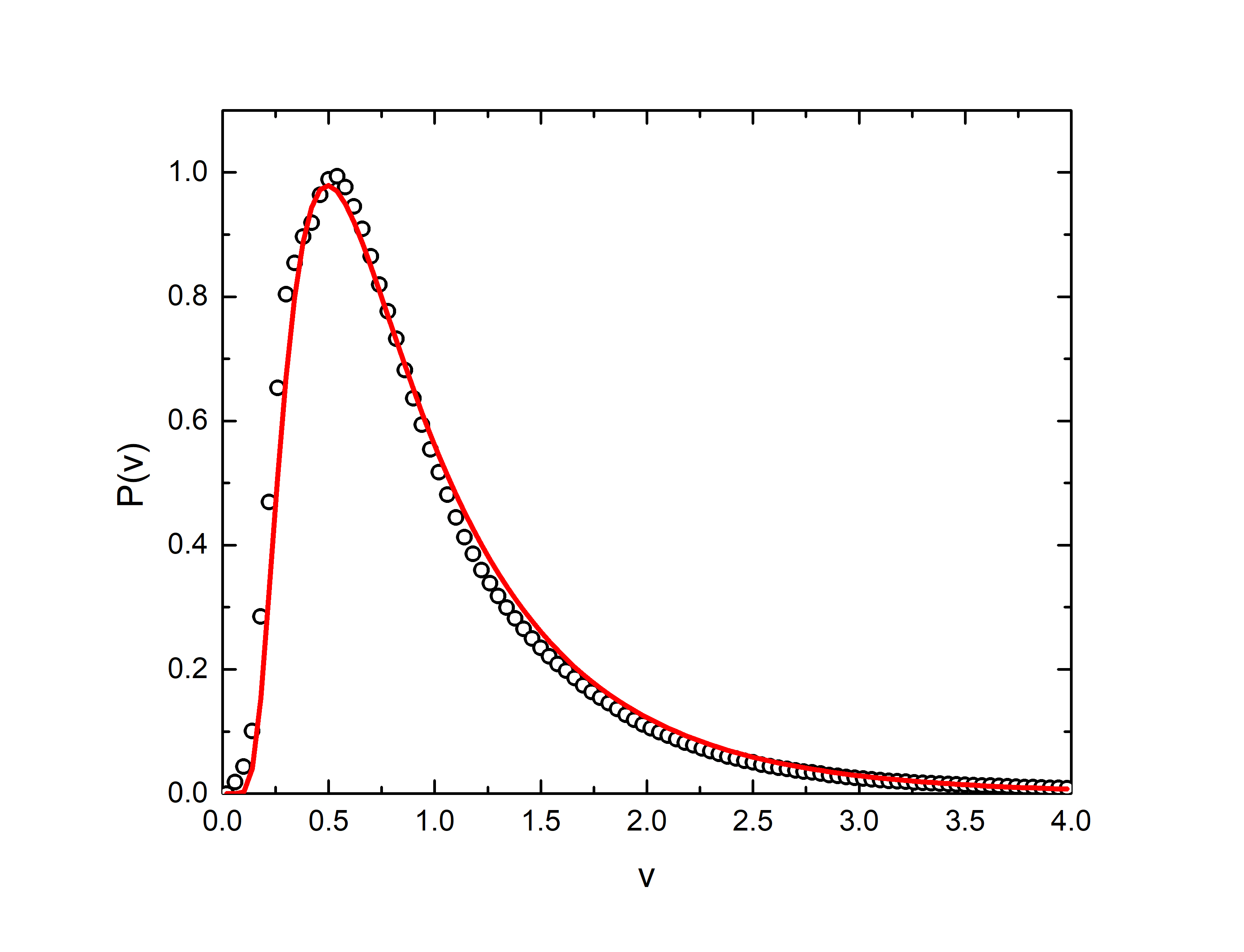}} 
		\caption{Experimental distribution $P(v)$ of the imaginary part 
			of the diagonal elements of the Wigner's $\hat K$
			matrix for the microwave undirected network $\Gamma$ at $\gamma =4.9 \pm 0.5$ (black open circles, frequency range $\nu \in [2,8]$ GHz). The theoretical
			distribution $P(v)$ evaluated from Eq.~(\ref{15}) for $\gamma =4.9$ is
			marked by red solid line. 
		}\label{Fig6}
	\end{center}
\end{figure}

\begin{figure}[tb]
\begin{center}
		\hspace*{-1.5cm} 
	\rotatebox{0}{\includegraphics[width=1.2\textwidth,
		height=1.2\textheight, keepaspectratio]{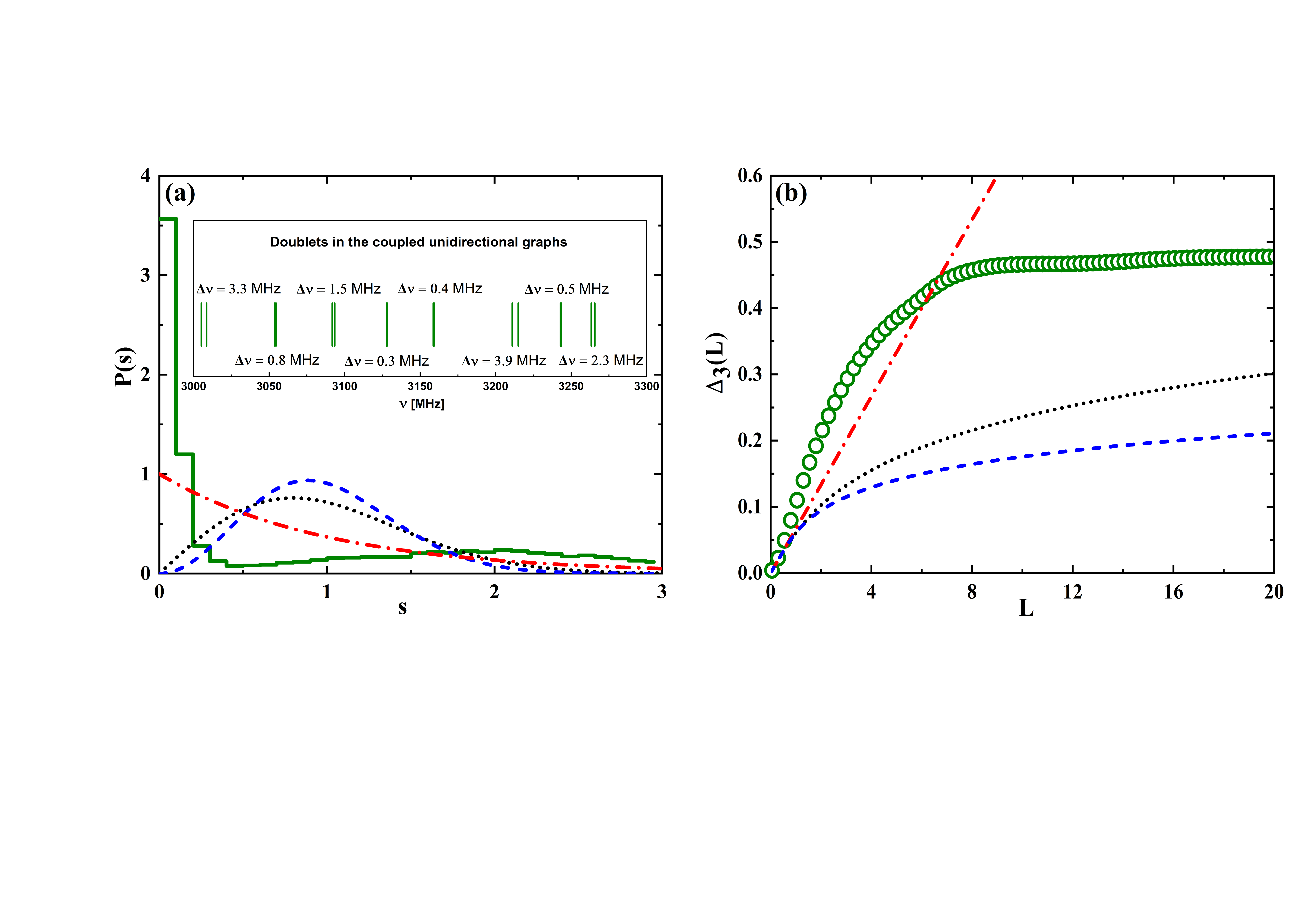}}
	\caption{(a) The numerical nearest-neighbor spacing distribution $P(s)$ (green histogram) and (b) the spectral rigidity $\Delta_3(L)$ (green open circles) calculated for the undirected graph $\Gamma$. The results in panels (a) and (b) are compared with the Poisson (red dotted-dashed line), GOE (black dotted line), and GUE (blue dashed line) predictions, respectively. The inset in the panel (a) shows an example of the spectrum of the graph $\Gamma$ (coupled unidirectional graphs $\Gamma_+$ and $\Gamma_-$) calculated in the frequency range $3.0-3.3$ GHz. The size of the doublets are specified in MHz.
	}\label{Fig7}
\end{center}
\end{figure}

\begin{figure}[tb] 
	\begin{center}
			\hspace*{-1.5cm}
		\rotatebox{0}{\includegraphics[width=1.2\textwidth,
			height=1.2\textheight, keepaspectratio]{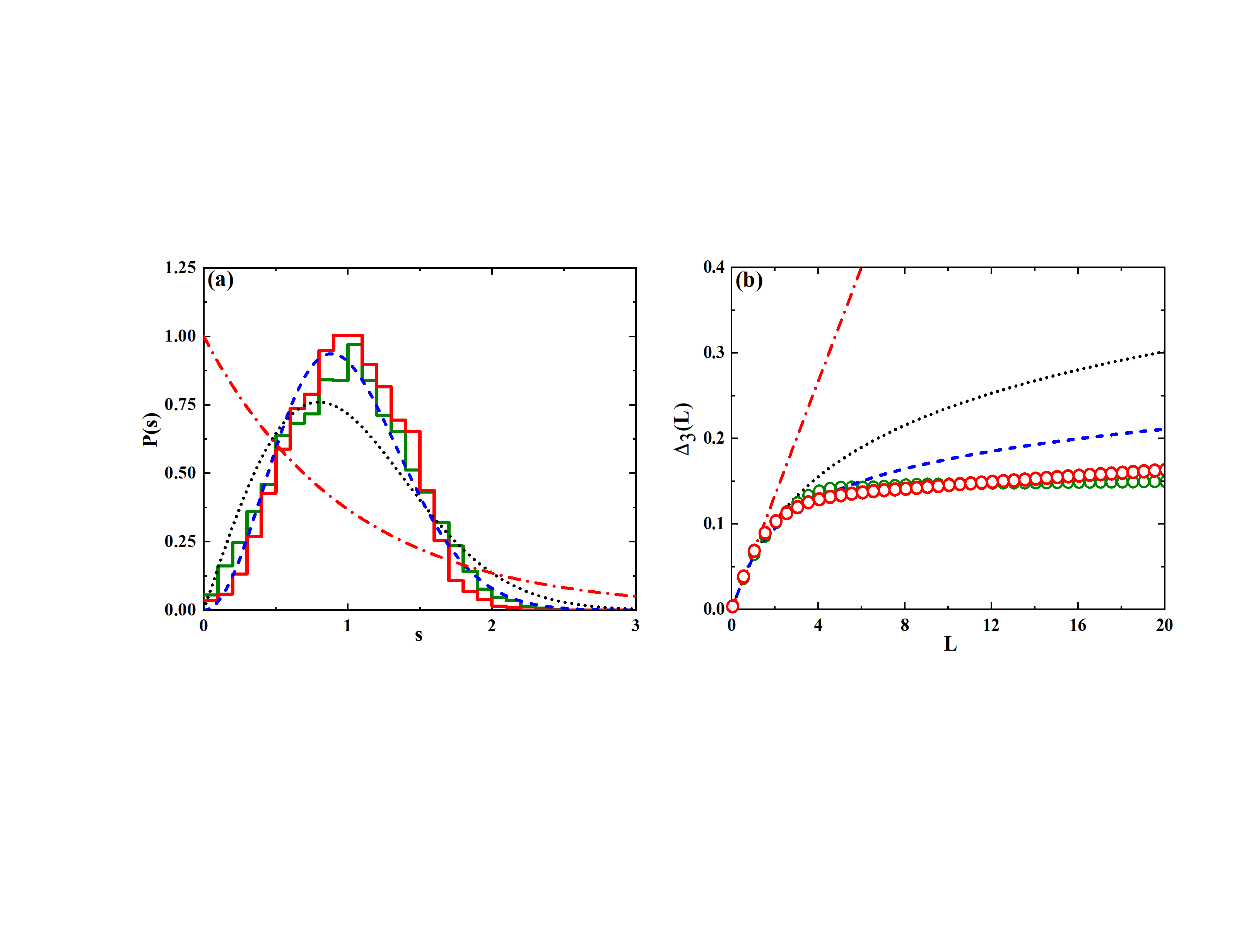}}
		\caption{(a) The numerical nearest-neighbor spacing distribution $P(s)$ evaluated from the spectra of the graph $\Gamma$ (see Fig.~1) under the assumption that the doublets are not resolved and are treated as singlet states (green histogram). The distribution $P(s)$ evaluated for the simplified closed graph $\Gamma'$ which was composed of 10 edges and 5 couplers (red histogram). The graph $\Gamma'$ is characterized by the spectrum of exactly doubly degenerate states. (b) The spectral rigidity $\Delta_3(L)$ evaluated from the spectra of the graph $\Gamma$ (green circles) is compared with the one evaluated for the graph $\Gamma'$ (red circles). The results presented in panels (a) and (b) are compared with the Poisson (red dotted-dashed line), GOE (black dotted line), and GUE (blue dashed line) predictions, respectively.	 
		}\label{Fig8}
	\end{center}
\end{figure}

\begin{figure}[tb]
	\begin{center}
		\rotatebox{0}{\includegraphics[width=0.8\textwidth,
			height=0.8\textheight, keepaspectratio]{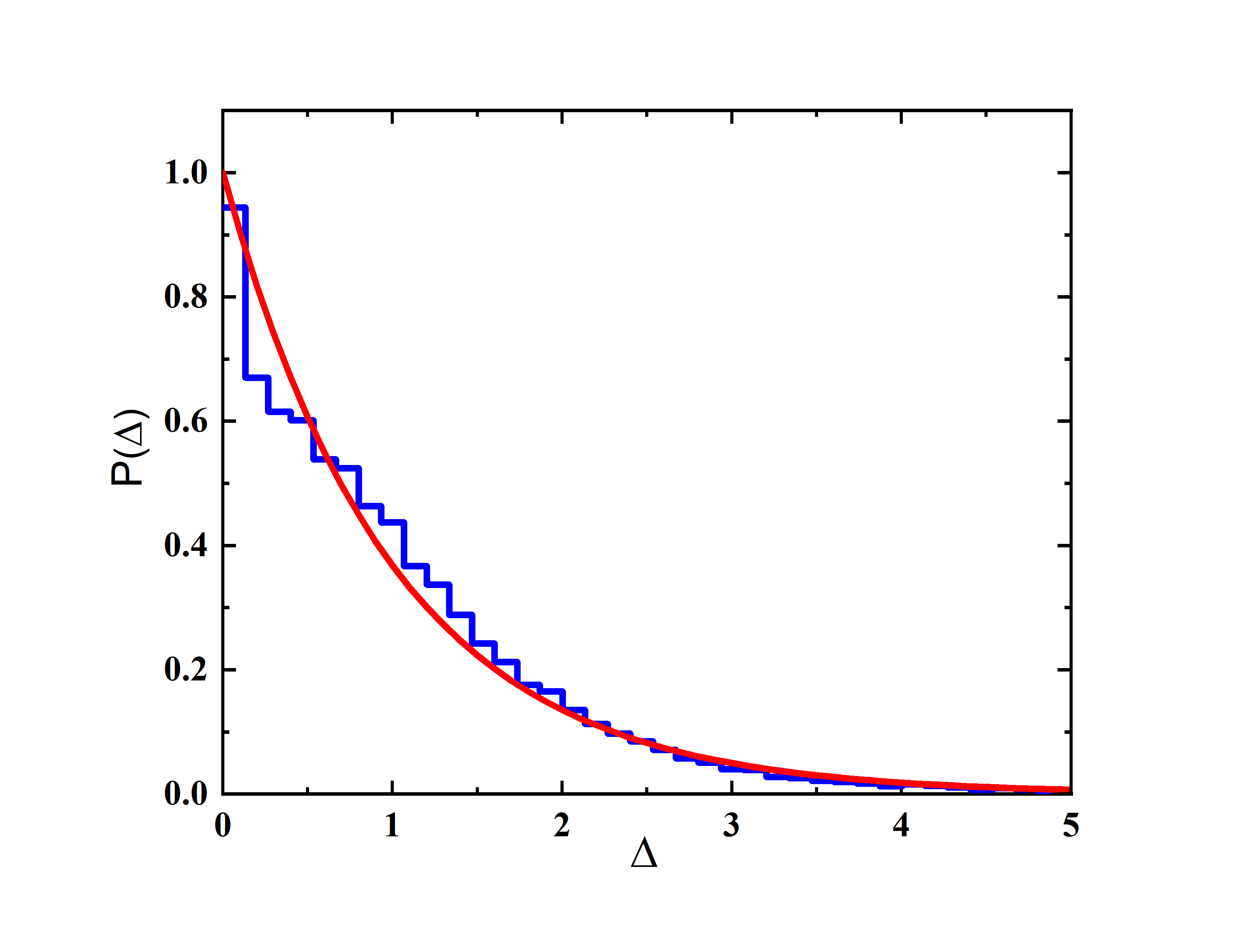}}
		\caption{The doublet size distribution $P(\Delta)$ of the graph $\Gamma$ (blue histogram) consisting of the coupled unidirectional graphs $\Gamma_{+} $ and $\Gamma_{-} $. The distribution $P(\Delta)$ is compared to the Poisson distribution $P_{Poisson} = \exp( -\Delta)$ (red solid line).	 
		}\label{Fig9}
	\end{center}
\end{figure}

\end{document}